\documentclass[12pt,tightenlines,eqsecnum,floats,aps,amsmath,amssymb,nofootinbib,prd,showpacs]{revtex4}

\usepackage{setspace}
\usepackage{amsmath,amssymb,amsfonts,amsthm}
\usepackage{graphicx}
\usepackage{enumerate}

\begin{document}
\title{Classical and quantum behavior of dynamical systems defined by functions of solvable Hamiltonians}

\author{J. Fernando Barbero G.}
\email[]{fbarbero@iem.cfmac.csic.es}
\affiliation{Instituto de
Estructura de la Materia, CSIC, Serrano 123, 28006 Madrid, Spain}

\author{I\~{n}aki Garay}
\email[]{igael@iem.cfmac.csic.es}
\affiliation{Instituto de
Estructura de la Materia, CSIC, Serrano 123, 28006 Madrid, Spain}

\author{Eduardo J. S. Villase\~{n}or}
\email[]{ejsanche@math.uc3m.es}
\affiliation{Grupo de
Modelizaci\'on y Simulaci\'on Num\'erica, Universidad Carlos III
de Madrid, Avda.\ de la Universidad 30, 28911 Legan\'es, Spain}
\affiliation{Instituto de Estructura de la Materia, CSIC, Serrano
123, 28006 Madrid, Spain}


\begin{abstract}
We discuss the classical and quantum mechanical evolution of systems
described by a Hamiltonian that is a function of a solvable one, both classically and quantum mechanically. The case in which the solvable Hamiltonian corresponds to
the harmonic oscillator is emphasized. We show that, in spite of the
similarities at the classical level, the quantum evolution is very different. In particular, this difference is important
in constructing coherent states, which is impossible in most cases.
The class of Hamiltonians we consider is interesting due to its
pedagogical value and its applicability to some open research problems
in quantum optics and quantum gravity.
\end{abstract}


\maketitle

\section{Introduction}

The goal of this article is to discuss the classical and quantum mechanics of
systems whose Hamiltonian $H=f(H_0)$ is a function of the
harmonic oscillator Hamiltonian $H_0$. The results can be
easily generalized to other choices of $H_0$ for which the classical and quantum equations of motion are exactly solvable.

Once we solve the classical equations of
motion for $H_0$, it is possible to study a
system described by $H=f(H_0)$. Although the solution is a straightforward
exercise in classical mechanics, we will discuss it in
detail because it is interesting to compare the solutions
corresponding to both classical and quantum dynamics. Quantum mechanically the
unitary evolution operator for $\hat{H}=f(\hat{H_0})$ can also be
constructed exactly once we know the evolution generated by
$\hat{H}_0$. A comparison of the dynamics given by
$\hat{H}_0$ and $f(\hat{H}_0)$ will allow us to analyze some distinctive features of the coherent
states of the harmonic oscillator and discuss the difficulties
that appear when we try to construct similar states for the
dynamics generated by $f(\hat{H_0})$. This comparison will help us understand
some aspects of the open problem of building appropriate
semiclassical states for general Hamiltonians.

Non-trivial systems whose evolution can be solved exactly both
classically and quantum mechanically are rare. Usually,
realistic systems are described by Hamiltonians of the form
$H=H_0+H_1$, where $H_0$ is a solvable Hamiltonian and $H_1$
represents a perturbation. In most cases it is impossible to give
the solutions to the equations defined $H$, so it is necessary to resort to approximation methods. The starting point of perturbation
theory is the known dynamics generated by $H_0$. The simplest choice of $H_0$ is the Hamiltonian of a free particle. However, if we
are considering a system that has bound states, it is much better
to consider a solvable $H_0$ with bound states, such as the harmonic
oscillator.

In this paper we consider a different way to perturb a solvable
Hamiltonian $H_0$ by considering a \textit{function} of it. If this function is
close to the identity, it will be possible to treat the system as
a perturbation of $H_0$ in the usual sense; otherwise it will provide
different but still solvable dynamics.

We point out that these kind of Hamiltonians
appear in some physical applications, for example, in the context
of quantum optics and classical and quantum gravity. For instance,
the propagation of light in Kerr media\cite{banerji,leonski}
--- media with a refractive index with a component that depends on
the intensity of the propagating electric field --- is described
(for a single mode field given by the creation and annihilation
operators $\hat{a}^\dagger$ and $\hat{a}$ and in the low loss
approximation) by
\begin{equation}
\hat{H}=\chi (\hat{N}^2-\hat{N}) =\chi \mbox{:$\hat{N}^2$:}=\chi\,
\hat{a}^{\dagger2} \hat{a}^2,
\end{equation}
where $\chi$ is related to the susceptibility of the medium, and
the Hamiltonian is a function of the number operator $\hat{N}=
\hat{a}^\dagger \hat{a}$. The symbol $:\,\,:$ denotes normal
ordering (creation operators to the left of the
annihilation ones) and the operators $\hat{a}$ and $\hat{a}^{\dagger}$ satisfy
the usual commutation relation $[\hat{a},\hat{a}^{\dagger}]=1$.

Another situation where we find this kind of Hamiltonian is in
general relativity. Einstein-Rosen\cite{Einstein}
waves are cylindrically symmetric solutions to the Einstein
equations in vacuum. The Hamiltonian that describes this system
is\cite{Ashtekar2,Fernando}
\begin{equation}
H=2(1-\exp(-H_0/2)), \label{ERham}
\end{equation}
where $H_0$ is a free (and easily solvable) Hamiltonian.

The examples we have mentioned are field theories with
Hamiltonians $f(H_0)$ that are functions of free Hamiltonians $H_0$
(i.e. quadratic in the fields and their canonical conjugate momenta)
describing an infinite number of harmonic oscillators. Although these models can be
solved, we will concentrate here on
finite dimensional examples to avoid field theoretical
complications, in particular, issues related to
the presence of an infinite number of degrees of freedom and the
coupling of the infinite different modes induced by the function $f$.

We consider
\begin{equation}
H=f(H_0), \qquad H_0=\frac{p^2}{2m}+\frac{k}{2}x^2.
\end{equation}
To simplify the calculations, we will assume
that $m=1$ and $k=1$. We will also work with an
arbitrary function $f$ (subject to some mild smoothness
conditions) until the very end of our discussion. At that point we will make some explicit calculations by
using the functional form of the Einstein-Rosen Hamiltonian. We emphasize that similar arguments could be made
for any system whose Hamiltonian is a function of a solvable one.

\section{Classical treatment}

We first discuss the classical solution for $H_0=\frac{1}{2}(p_0^2+x_0^2)$. The dynamics generated by $H_0$ is given by the
equations
\begin{equation}
\frac{dx_0}{dT}=p_0,\qquad \frac{dp_0}{dT}=-x_0.
\end{equation}
Here we denote the time parameter as $T$ because in the following we will compare two types of related dynamics where two different time parameters will be relevant. The general solution to these equations can be written as:
\begin{subequations}
\label{5}
\begin{align}
x_0(T)&=\frac{1}{\sqrt{2}}(ae^{-iT}+\bar{a}e^{iT}),\\
p_0(T)&=\frac{-i}{\sqrt{2}}(ae^{-iT}-\bar{a}e^{iT}),
\end{align}
\end{subequations}
where $a$ and its complex conjugate, denoted as $\bar{a}$, are fixed by the
initial conditions (at $T=0$)
\begin{eqnarray}
a=\frac{x_0+ip_0}{\sqrt{2}}\,.
\end{eqnarray}
In view of this last expression it is useful to introduce a complex variable $z_0=x_0+ip_0$ to describe positions and momenta simultaneously. In particular Eq.~(\ref{5}) can be rewritten as
\begin{equation}
z_0(T)=z_0e^{-iT}.
\end{equation}
The trajectories in phase space, described now as the complex $z$-plane, are circumferences centered in the origin with radius $|z_0|=\sqrt{x_0^2+p_0^2}=\sqrt{2H_0}$.

Consider next the solutions for $H=f(H_0)$. To have well-defined equations of motion we require that the function $f$ be
differentiable. The equations of motion now read
\begin{subequations}
\label{class}
\begin{align}
\frac{dx}{dt}&=\{x,f(H_0)\}=f^\prime(H_0)p,\label{class1}\\
\frac{dp}{dt}&=\{p,f(H_0)\}=-f^\prime(H_0)x,\label{class2}
\end{align}
\end{subequations}
where $f^\prime$ denotes the derivative of $f$ with respect to its
argument and $\{\,\,,\,\}$ is the Poisson bracket. In principle,
these coupled, non-linear, differential equations might seem difficult
to solve, but because $H_0$ is a constant of motion,
\begin{equation}
\frac{dH_0}{dt}=p\frac{dp}{dt}+x\frac{dx}{dt}=-f^\prime(H_0)px+f^\prime(H_0)xp=0,
\end{equation}
we can simplify them by introducing a new time parameter
\begin{equation}
T(t)=f^\prime(H_0)t.
\label{reparam}
\end{equation}
The reparametrization given by Eq.~(\ref{reparam}) allows us to transform Eq.~(\ref{class}) into the form of Eq.~\eqref{5}
corresponding to the harmonic oscillator with the solution
\begin{subequations}
\label{xp}
\begin{align}
x(t)&=x_0(T(t))=\frac{1}{\sqrt{2}}\big[ae^{-if^\prime(H_0)t}+\bar{a}e^{if^\prime(H_0)t}\big],\label{x}\\
p(t)&=p_0(T(t))=\frac{-i}{\sqrt{2}}\big[ae^{-if^\prime(H_0)t}-\bar{a}e^{if^\prime(H_0)t}\big].\label{p}
\end{align}
\end{subequations}
Note that although $(x(t),p(t))$ have the same
physical meaning as $(x_0(T),p_0(T))$, they are different functions
-- $x(t)$ is the composition (in the mathematical sense) $x_0(T(t))$ of $x_0(T)$ and $T(t)$. We find
in Eq.~(\ref{xp}) an energy dependent definition of time
that yields a different time evolution for each solution to the
equations of motion. ($H_0$ has a different value for each initial
condition.) The orbits in phase space for
$H_0$ and $f(H_0)$ are the same taken as non-parametrized
curves. Nevertheless, for $H_0$ the curves are
parametrized by $T$, whereas for $f(H_0)$ they are parametrized by
$t$. The solutions for $H_0$ all have the same frequency
\begin{equation}
z_0(T)=z_0e^{-iT}\,,
\end{equation}
in contrast to those for $f(H_0)$ which have frequencies that depend on the
initial conditions (through the value of $H_0=|z_0|^2/2$)
\begin{equation}
z(t)=z_0e^{-itf^\prime(H_0)}\label{13}.
\end{equation}

\section{quantum evolution}

The behavior of quantum systems is quite different from the
classical one. We choose as a basis for the Hilbert space of
the harmonic oscillator the states $|n\rangle$ which satisfy
$\hat{H}_0|n\rangle=\hbar(n+1/2)|n\rangle$. (In the following we
choose units such that $\hbar=1$.) Every initial state can be
expressed as
\begin{equation}
|\psi\rangle=\sum_{n=0}^\infty c_n|n\rangle,
\end{equation}
and the evolution is given by:
\begin{equation}
|\Psi_0(T)\rangle=e^{-i\hat{H}_0(T-T_0)}|\psi\rangle=\sum_{n=0}^\infty
c_ne^{-i(T-T_0)(n+1/2)}|n\rangle.
\end{equation}
Let us consider a Hamiltonian defined as
$\hat{H}=f(\hat{H_0})$. To define $f(\hat{A})$ for a
general self-adjoint operator $\hat{A}$ we must require that $f$
satisfy the relevant conditions for the spectral theorems.
\cite{reedsimon} In our case any function defined on the spectrum
of $\hat{H}_0$ would give rise to a well defined Hamiltonian, but
because we want to discuss the semiclassical limit, we will
require that $f$ be differentiable.

The eigenvectors $|n\rangle$
of the Hamiltonian $\hat{H_0}$ are also eigenvectors of
$f(\hat{H_0})$ with eigenvalues $f(n+1/2)$. The
evolution of a state $|\psi\rangle$ defined by $f(\hat{H_0})$ is
given by
\begin{equation}
|\Psi(t)\rangle=e^{-if(\hat{H}_0)(t-t_0)}|\psi\rangle=
\sum_{n=0}^{\infty}c_ne^{-i(t-t_0)f(n+1/2)}
|n\rangle.\label{evolution}
\end{equation}
We see in Eq.~(\ref{evolution}) that the situation is
not analogous to that found in the classical system. In the
quantum mechanical case we cannot obtain $|\Psi(t)\rangle$ from
$|\Psi_0(T)\rangle$ by a simple {reparametrization of
time, even if we allow it to depend on the initial state vector
$|\psi\rangle$, because the relative phases between different
energy eigenstates $|n\rangle$ change in time and produce a non-trivial difference between the wave
functions under the evolution defined by $\hat{H}_0$ and
$f(\hat{H}_0)$.

\section{Coherent states}

Once we know the exact classical evolution of any state, we can search for
semiclassical states that evolve in the same way. In general, even for the harmonic oscillator, wave packets
(more specifically their squared modulus) change shape as they evolve in time.\cite{galindo,messiah} However, there is a family of
non-stationary coherent states whose wave function $\psi$
(modulus squared) does not change its shape as time evolves.
A plot of $|\psi|^2$ as a function of time shows that it rigidly
moves back and forth as a particle subject to a restoring force
proportional to the distance to a fixed point in space, that is, a
classical harmonic oscillator with Hamiltonian
$H_0=\frac{1}{2}(p^2+x^2)$ (see Fig.~\ref{coherent}).

\begin{figure}[h]
\includegraphics[width=.45\textwidth]{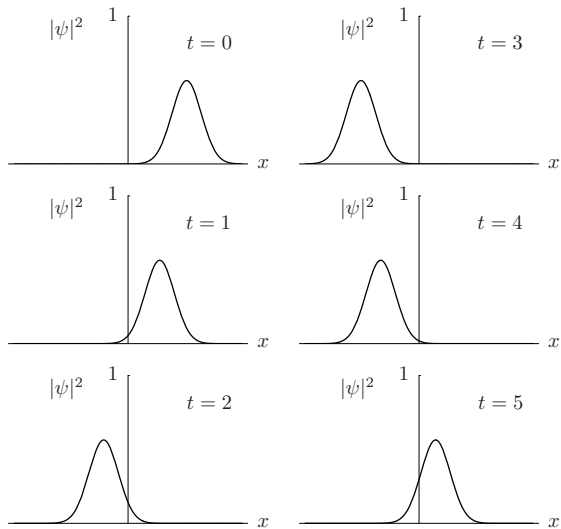}
\caption{Evolution of the squared modulus of the wave function of
a coherent state for the harmonic oscillator.} \label{coherent}
\end{figure}

These coherent states of the harmonic oscillator (and their free
field counterparts) have a number of additional interesting
properties including the following\cite{galindo}:

\begin{enumerate}

\item They are eigenstates of the annihilation operator --that can be written in terms of the position and momentum operators as
$\hat{a}=\frac{1}{\sqrt{2}}(\hat{X}+i\hat{P})$--  with \textit{complex}
eigenvalue $z$ whose real and imaginary parts encode the initial position and momenta of the
classical motion. In terms of $z$ and its complex conjugate $\bar{z}$ we have $\langle
z|\hat{X}|z\rangle=\frac{1}{\sqrt{2}}(\bar{z}+z)$ and $\langle
z|\hat{P}|z\rangle=\frac{i}{\sqrt{2}}(\bar{z}-z)$. If we start with the condition that $|z\rangle$ is an eigenstate
of $\hat{a}$, it is straightforward to express $|z\rangle$ in terms of the
energy eigenstates $|n\rangle$:
\begin{equation}
|z\rangle=\exp(-|z|^2/2)\sum_{n=0}^\infty\frac{z^n}{\sqrt{n!}}|n\rangle.
\end{equation}

\item The dispersion of the position and momentum
operators in these states $[\langle z|\hat{X}^2|z\rangle-\langle
z|\hat{X}|z\rangle^2] = [\langle
z|\hat{P}^2|z\rangle-\langle z|\hat{P}|z\rangle^2]=\frac{1}{2}$ are
constant and saturate the Heisenberg uncertainty inequalities
(coherent states define minimal wave packets). It
can be seen that coherent states are also minimal with respect to
energy and momentum \cite{galindo} in the sense that $\Delta_zH_0=|z|$, the characteristic time of the system in the state $|z\rangle$ is $\tau_z=1/(2\Delta_zH_0)$, and then $\tau_z\Delta_zH_0=1/2$.

\item The time evolution of the state $|z_0\rangle$ is
given by
\begin{equation}
e^{-iT\hat{H}_0}|z_0\rangle=e^{-i T/2}| e^{-i T} z_0\rangle.
\label{evolz}
\end{equation}
This equation means that as time evolves, the unitary ray
defined by a coherent state $|z_0\rangle$ at $T=0$ (i.e. the set of vectors of the form $e^{i\theta}|z_0\rangle$ with $\theta\in\mathbb{R}$) remains coherent at any time $T$ and is labeled by
\begin{equation}
z(T)=e^{-iT}z_0=x_0(T)+ip_0(T),
\end{equation}
where the functions $x_0(T)$ and $p_0(T)$, given by Eq.~(\ref{5}), are the position and momentum of the classical harmonic oscillator as a function of time.

\item The set of coherent states defines a linear, non-orthonormal, overcomplete
basis of the Hilbert space for a harmonic oscillator. In particular,
we can find a relation of the type
\begin{equation}
\frac{1}{\pi}\int_{\mathbb{C}}\mathrm{d}^2z|z\rangle\langle
z|=\mathbb{I}.
\end{equation}
As we can see, the coherent states for the harmonic oscillator satisfy a
set of properties that allow us to consider them as
semiclassical in the sense that their time evolution closely follows
the classical one. They also satisfy interesting
properties that render them an important tool in the study of
oscillator systems or free field theories.

\end{enumerate}

\section{Example}

As an illustration of these methods, we will answer the question: Can we
build appropriate coherent states for a one-particle system with a
Hamiltonian of the form} $f(H_0)$ with
$H_0=\frac{1}{2}(p^2+x^2)$? This case is interesting
because if the answer were affirmative,
it could be possible to extend the result for interesting field
theories such as general relativity reductions of the
Einstein-Rosen type. As we will see the answer to this question is in the negative.

We show that it is not possible to build proper coherent states for $f(H_0)$ by proving that under time evolution the label $z$, which
encodes the initial data, cannot evolve according to the
classical dynamics dictated by $f(H_0)$. In terms of the
initial data $z_0=x_0+i p_0$ (combined in the complex number $z_0$), the classical evolution of the
system is obtained from Eq.~(\ref{13})
\begin{equation}
z(t)=z_0\exp\big(-it f^{\prime}(|z_0|^2/2)\big). \label{evolfz}
\end{equation}
So we will require that the state, which we will also label
$|z\rangle$ in analogy with the usual coherent states, evolve as
\begin{equation}
\exp(-itf(\hat{H_0}))|z\rangle=
\exp(i\varphi(t,z))\left|z\exp(-itf^{\prime}(|z|^2/2))
\right\rangle. \label{evolfpsiz}
\end{equation}
Equation~\eqref{evolfpsiz} is similar to Eq.~(\ref{evolz}). Note that we must work with unitary rays so we
include an arbitrary phase $\exp(i\varphi(t,z))$. We
now expand $|z\rangle$ in the orthonormal basis provided by the
energy eigenfunctions $|n\rangle$ of the harmonic oscillator
Hamiltonian
\begin{equation}
|z\rangle=\sum_{n=0}^{\infty}\psi_n(z)|n\rangle,
\end{equation}
where the coefficients $\psi_n(z)=\psi_n(x+ip) =\psi_n(x,p)$ are
taken as differentiable functions. Equation~(\ref{evolfpsiz}) becomes
\begin{equation}
\psi_n(z)=\exp[i(\varphi(t,z)+tf(E_n))]\psi_n(z\exp(-itf^{\prime}(|z|^2/2))).
\label{evolfpsiz1}
\end{equation}
If we use the notation $\psi_n(x,p)$ for $\psi_n(z)$ we can rewrite Eq.~(\ref{evolfpsiz1})
as
\begin{eqnarray}
\psi_n(x,p)&=& \exp[i(\varphi(t,z)+t
f(E_n))] \label{eq:last}\\
&&\hspace{-8mm}\times\psi_n\Big(x\cos(tf^{\prime}(|z|^2/2)) +
p\sin(tf^{\prime}(|z|^2/2)), p\cos(tf^{\prime}(|z|^2/2))-x\sin(tf^{\prime}(|z|^2/2))\Big). 
\nonumber
\end{eqnarray}
The left-hand side of Eq.~\eqref{eq:last} does not depend on time,
so the time derivative of the right-hand side must be zero. We evaluate this derivative at $t=0$ and obtain
the consistency condition
\begin{equation}
f^{\prime}(|z|^2/2)\left[p\frac{\partial\psi_n}{\partial x}(x,p) -
x\frac{\partial\psi_n}{\partial p}(x,p)\right] = -
i\left[\dot{\varphi}_0(z)+f(E_n)\right]\psi_n(x,p),\label{consist}
\end{equation}
with $\dot{\varphi}_0(z)=\frac{\partial\varphi(0,z)}{\partial t}$. By
introducing polar coordinates $x=r\cos\theta$ and $p=r\sin\theta$ we
can rewrite Eq.~(\ref{consist}) as
\begin{equation}
\frac{\partial\psi_n(r,\theta)}{\partial\theta}=i\frac{\dot{\varphi}_0(r,\theta)
+f(E_n)}{f^{\prime}(r^2/2)}\,\psi_n(r,\theta). \label{consist1}
\end{equation}
Equation~\eqref{consist1} can be solved to give
\begin{equation}
\psi_n(r,\theta)=c_n(r^2) \exp\left[i\frac{f(E_n)\theta
+X(\theta,r)}{f^{\prime}(r^2/2)}\right], \label{solution}
\end{equation}
where
\begin{equation}
X(\theta,r) =
\!\int\dot{\varphi}_0(\theta,r)\,\mathrm{d}\theta,
\end{equation}
and $c_n(r^2)$ are
arbitrary functions of $r^2$. It can be easily checked that
for the usual harmonic oscillator, $f(x)=x$ and $E_n=n+\frac{1}{2}$,
the choice $\varphi(t,z)=-t/2$ gives
$\psi_n=c_n(r^2)\exp(in\theta)$. The latter can be written as
$\psi_n(z)=c_n(|z|^2)\exp(in \mathrm{Arg}_n(z))$, where
$\mathrm{Arg}_n$ is a branch of the argument of $z$. With this
choice $\exp(in \mathrm{Arg}_n(z))$ is independent of the branch
chosen for the argument, and we can write
$\psi_n(z)=c_n(|z|^2)\exp(in \mathrm{Arg}(z))=\tilde{c}_n(|z|^2)z^n$
with $\mathrm{Arg}(z)=\{\arctan(p/q)+2k\pi,k\in\mathbb{Z}\}$. This
result should be compared with the result
$\psi_n(z)=e^{-|z|^2/2}z^n/n!$ corresponding to the harmonic
oscillator coherent states. As we can see only part of the
dependence on $z$ is fixed by Eq.~\eqref{evolfpsiz}, but the result is
compatible with $\psi_n(z)=e^{-|z|^2/2}z^n/n!$. By using the other
conditions the full dependence on $z$ can
be obtained.

From Eq.~(\ref{solution}) we observe that, in general, the
result will depend on the branch chosen. This ambiguity is
unacceptable so we conclude that it is usually impossible
to have a family of coherent states satisfying a condition
equivalent to Eq.~\eqref{evolfpsiz} for the evolution given by $f(H_0)$.
We consider an explicit example using the functional form
given by the Hamiltonian of the Einstein-Rosen waves
$f(x)=2(1-e^{-x/2})$. The solution (\ref{solution}) for this
choice is
\begin{equation}
\psi_n(r,\theta)=c_n(r^2)
\exp\left(ie^{r^2/4}[2(1-e^{-\frac{1}{2}(n+\frac{1}{2})})\theta+X(\theta,r)]\right).
\label{solutionf}
\end{equation}
We need to require that
\begin{equation}
\exp\left(ie^{r^2/4}[2(1-e^{-\frac{1}{2}(n+\frac{1}{2})})\theta+X(\theta,r)]\right)
\end{equation}
be independent of the branch chosen for the argument $\theta$ --otherwise it is not single-valued. However, this requirement is impossible to satisfy because $X(\theta,r)$
is independent of $n$. If we write
$\theta=\tilde{\theta}+2k\pi$, we obtain the
condition
\begin{equation}
\label{eq:last2}
4k\pi
e^{r^2/4}(1-e^{-\frac{1}{2}(n+\frac{1}{2})})+e^{r^2/4}X(\tilde{\theta}+2k\pi,r)\equiv
e^{r^2/4}X(\tilde{\theta},r) \quad (\mathrm{mod}\, 2\pi).
\end{equation}
If we consider Eq.~\eqref{eq:last2} for two different numbers $n$
and $m$, the difference between them gives
\begin{equation}
4\pi k\,e^{r^2/4}\left(e^{-\frac{1}{2}(n+\frac{1}{2})}
-e^{-\frac{1}{2}(m+\frac{1}{2})}\right)\equiv0\quad (\mathrm{mod}\,
2\pi)
\end{equation}
for all $m$ and $n$ which is
impossible.

\section{Conclusions}

We have described the classical and
quantum dynamics of systems with Hamiltonians that are functions
of other solvable Hamiltonians and compared the cases $H_0$ and $f(H_0)$. Classically the states
evolve in very similar ways and follow the same phase space orbits
although with different time parametrizations. In contrast,
their quantum evolution is qualitatively different
due to the appearance of non-trivial relative phases. We
discussed this issue by analyzing the existence of coherent states and
their properties for functionally related Hamiltonians.
In particular, we gave a proof of the impossibility of
constructing coherent states that satisfy the four conditions in Sec.~IV for general Hamiltonians of the form
$H=f(H_0)$, with $H_0$ corresponding to the harmonic oscillator.
This case is especially
significant because of the role played by harmonic oscillators in the
description of free quantum field theories.

By relaxing some of the conditions defining coherent
states for the harmonic oscillator, we can conceivably find a set
of suitable semiclassical states for the more complicated dynamics
given by $f(H_0)$. Our analysis does not exclude this possibility,
but suggests that the definition that we must
adopt will require major changes in the conditions that are satisfied by the
familiar coherent states.

\begin{acknowledgments}
We want to thank A.\ Ashtekar, G.\ Mena Marug\'an, and M.\
Varadarajan for discussions on this issue. We also thank the
referees for their thorough revision of the manuscript and
thoughtful comments. I\~{n}aki Garay is supported by a Spanish
Ministry of Science and Education under the FPU program. This work
is also supported by the Spanish MEC under the research grant
FIS2005-05736-C03-02.
\end{acknowledgments}



\begin{thebibliography}{8}

\bibitem{banerji} J. Banerji, ``Nonlinear wave packet dynamics of coherent states,'' Pramana J. Phys. {\bf 56}, 267--280 (2001).

\bibitem{leonski} W. Leo\'{n}ski, ``Periodic behaviour of displaced Kerr states,'' Acta Phys. Slovaca {\bf 48}, 371--378 (1998).

\bibitem{Einstein} A. Einstein and N. Rosen, ``On gravitational waves,'' J. Franklin Inst. {\bf 223}, 43--54 (1937).

\bibitem{Ashtekar2} A. Ashtekar and M. Varadarajan, ``Striking property of the gravitational Hamiltonian,'' Phys. Rev. {\bf D} 50, 4944--4956 (1994).

\bibitem{Fernando} J. F. Barbero G., I. Garay, and E. J. S. Villase\~{n}or, ``Probing quantized Einstein-Rosen waves with massless scalar matter,'' Phys. Rev. D {\bf 74}, 044004-1--22 (2006).

\bibitem{reedsimon} M. Reed and B. Simon, \textit{Methods of Modern Mathematical Physics: Functional Analysis} (Academic Press, London 1980),
Vol. 1.

\bibitem{galindo} A. Galindo and P. Pascual, \textit{Quantum Mechanics} (Springer-Verlag, Berlin 1991), Vols. I and II.

\bibitem{messiah} A. Messiah, \textit{Quantum Mechanics} (Dover, N.Y. 1999).

\end{thebibliography}
\end{document}